# Stellar Polarimetry:
# Where Are We and Where Are We Going?

Jennifer L. Hoffman, John C. Brown, Kenneth Nordsieck,
Nicole St-Louis, and Gregg Wade

# Stellar Polarimetry:
# Where Are We and Where Are We Going?

Jennifer L. Hoffman[a], John C. Brown[b], Kenneth Nordsieck[c], Nicole St-Louis[d], and Gregg Wade[e]

[a]Department of Physics & Astronomy, University of Denver, 2112 E. Wesley Ave., Denver, CO 80208-6900, USA
[b]School of Physics and Astronomy, University of Glasgow, G12 8QQ, UK
[c]Department of Astronomy, University of Wisconsin-Madison, 475 N. Charter Street, Madison, WI, 53706-158,. USA
[d] Université de Montréal & Centre de Recherche en Astrophysique du Québec (CRAQ), C.P. 6128, Succ. Centre Ville, Montréal (Qc), H3C 3J7, Canada
[e]Department of Physics, Royal Military College of Canada, PO Box 17000, Stn Forces, Kingston, Ontario K7K 7B4, Canada

**Abstract.** On the final day of the Stellar Polarimetry conference, participants split up into three "breakout sessions" to discuss the future of the field in the areas of instrumentation, upcoming opportunities, and community priorities. This contribution compiles the major recommendations arising from each breakout session. We hope that the polarimetric community will find these ideas useful as we consider how to maintain the vitality of polarimetry in the coming years.

**Keywords:** instrumentation: polarimeters; techniques: polarimetric
**PACS:** 95.30.Gv, 95.55.Qf, 95.75.Hi; 97.10Ld

## INTRODUCTION

The conference organizers felt strongly that the Stellar Polarimetry meeting should be a true "workshop," during which members of the stellar polarimetry community could learn from one another and make progress on common goals. In order to do this, they ended each of the first three meeting days with a moderated discussion period; they also reserved the last morning for small-group "breakout sessions" and a closing panel session at which the breakout session chairs presented the results from their groups and led a final discussion.

In this contribution, the members of the panel summarize the recommendations and ideas that arose from the three breakout sessions, which focused on instrumentation (led by Kenneth Nordsieck), upcoming opportunities (led by Gregg Wade and Nicole St-Louis), and community priorities (led by Jennifer Hoffman). We were tasked by the conference organizers to use these discussions to encourage active stellar polarimetry researchers to reflect on the current status of the field as well as the directions they would like to see it take in the near future.

We have not attempted to transcribe the extensive and productive conversations that took place in response to the panel reports, but rather summarize the results presented by each breakout session chair along with ideas and suggestions that arose in response to them during the closing discussion. We encourage everyone with an interest in stellar polarimetry to take advantage of opportunities to become involved in shaping the future of the field; we hope this document will serve as a record of the community's needs and priorities.

## Breakout Session A: Instrumentation

This session was chaired by Kenneth Nordsieck (University of Wisconsin-Madison). Session participants focused on two main topics: highest priorities for new instrumentation and recommendations for "what to do" and "what not to do" when designing and building a polarimetric instrument.

Highest instrumentation priorities identified by the panel and by informal discussions at the meeting were spectropolarimeters, polarimeters with very high resolution, and a polarimetric all-sky survey instrument. Existing coverage could be improved in three domains: in time with high-speed instruments and regular monitoring programs; in space by adaptive optics and larger fields of view; and in wavelength by development of new instruments with sensitivities in the UV, X-ray, infrared, and near-infrared.

The session participants had several suggestions for those involved in future polarimetric instrument development. They felt it was particularly important to build polarimetric optics into the instrument design from the beginning, as "add-ons" tend to be more difficult and less reliable (though there are examples of successful add-on polarimeters). The community has a role to play in ensuring that these polarimetric capabilities remain in the final instrument design; to do this, we need to be vocal about the demand for these capabilities and their scientific importance, a theme that arose frequently in the overall panel discussion.

Awareness of limits to precision is important in any new instrument design, but community participation can also help with this by sharing "tricks learned by experience" among groups. For example, issues related to guiding, flat-fielding, and second-order optical effects may be more easily addressed if there is a mechanism for sharing expertise. Session participants recommended establishing a wiki devoted to polarimetric instrumentation where this information could be made widely available. Further discussion explored the idea of a "super-wiki" that could link to existing information and brought up the suggestion of an American Astronomical Society-sponsored "landing page" for information about astronomical polarimetry.

Much future polarimetric instrumentation will necessarily be cross-disciplinary in nature. The session participants discussed the difficulties in coupling polarimetry with adaptive optics, which leads to problems establishing polarimetric capabilities on extremely large telescopes (ELTs). In addition, difficulties in coupling polarimetry with cryogenic instruments can lead to problems developing NIR and IR spectropolarimeters. Audience members pointed out that an IR polarimeter is among the next-generation instruments proposed for SOFIA and suggested that the community express its support for this project. A polarimetry mailing list along the lines of the Massive Star Newsletter might be a productive way to organize interested community members and build support for these instruments as opportunities arise.

Finally, the session participants focused on Gemini GPOL as a particularly important upcoming instrument and suggested two strategies that could help ensure that it gets commissioned. First, assembling information on community demand for the ESO FORS polarimeters would help build science cases and argue that GPOL would be widely used. Second, a survey or questionnaire could demonstrate the need and desire for such an instrument among active researchers. Audience members suggested that those involved with Gemini should keep bringing up the topic of polarimetric instrumentation at future meetings to try to keep the issue at the forefront.

## Breakout Session B: Upcoming Opportunities

This session was chaired by Gregg Wade (Royal Military College of Canada) and presented at the panel discussion by Nicole St-Louis (Université de Montréal). Participants focused on three areas: important upcoming science goals, observational platforms to address these science goals, and modeling needs.

The question of a compelling polarimetric science goal came up repeatedly during the conference, but there was no consensus about one particular area that could galvanize support in the broader astronomical community. The upside to this is that there is a wide variety of interesting polarimetric science goals, including (but not, of course, limited to) jets from young stars, clumps in hot-star winds, low-level stellar magnetic fields, monitoring of variables, SN and GRB follow-up, exoplanet atmospheres, low-metallicity effects in chemical evolution, Galactic structure, and high-resolution imaging polarimetry.

In discussing observational platforms to address new science goals, the session participants separated telescope facilities by aperture. The next generation of ELTs should offer about 100 times the flux of 4m-class telescopes, which translates into an increase of 5 mag in detectable brightness or a 10x greater signal-to-noise ratio at a given magnitude. For ELT science, participants estimated we would need spectral resolving powers greater than 10,000, but were pessimistic about the possibilities of getting any polarimeter onto an ELT. To do so, they felt we would need to present a single compelling science case with broad interest beyond the field of stellar polarimetry, and this proved difficult to identify. State-of-the-art polarimetric instruments exist on several 8–10m-class telescopes, including Subaru (FOCAS), VLT (FORS2, NACO), Keck (LRIS), and Gemini (Michelle, GPI). However, this list is short and many are visiting instruments or soon to be decommissioned. The best chance for future opportunities may exist on smaller telescopes, less than 4 m in diameter. Polarimeters on such telescopes could be used for monitoring, rapid-response follow-up, or all-sky surveys. SOUTH POL is an example of such a planned survey (see [1]); perhaps this is a good argument for a similar survey in the northern hemisphere. Session participants also suggested that a smaller telescope about to be closed down might provide a good home for a dedicated polarimeter.

In this session, several modeling needs were identified. These included realistic, NLTE polarimetric radiative transfer codes in all four Stokes parameters, Sobolev exact integration methods for polarized P Cygni line profiles, and detailed hydrodynamic modeling. In order to develop these, more high-powered computational facilities are needed. These needs notwithstanding, members of the audience pointed out that simple analytical models that do not require advanced computing methods or clusters still have a strong role to play in polarimetric analysis.

## Breakout Session C: Community Priorities

This session was chaired by Jennifer Hoffman (University of Denver). The session began with a discussion of ideas that had been brought up on the first day of the conference, particularly those focusing on broadening the use of polarimetry in astronomy. Like the participants in the instrumentation session, this group suggested establishing a polarimetry-focused blog, wiki, or mailing list to bring together conference participants and others interested in polarimetry. Such a wiki might potentially be hosted at the AstroBetter website (http://www.astrobetter.com/wiki), which maintains a list of informational pages on various astronomical topics and methods that can be edited by registered participants. The attendee list for this conference could serve as a starting point around which to build a larger community mailing list. Several participants felt that establishing a common data reduction pipeline and collections of polarimetric standard stars would encourage a broader use of polarimetry in astronomy, since the technique is widely considered to be intimidatingly difficult. An online repository could also include tutorials to educate students and novices in polarimetric data reduction and interpretation and pre-developed lesson plans aimed at undergraduate astronomy majors.

Another idea for broadening participation was to develop a summer school or workshop to train new students and researchers in polarimetric methods; perhaps we could begin by contributing a single day or half-day session to an existing school. A special journal issue, series of review articles (e.g., in *Astrophysics and Space Science*), or book on polarimetric techniques might also fill a needed niche. Existing proposals and white papers (e.g., [2, 3, 4, 5, 6, 7, 8, 9]) could form the basis of a more formal publication. Session participants also felt that conference proceedings could play a key role in establishing a polarimetric literature; to that end, they suggested that the next polarimetry conference should include more review talks that would form useful introduction to the topic for the general astronomical public. In addition, a planned workshop on polarimetric opportunities for the E-ELT could galvanize the community and pave the way for polarimetry on the next generation of large telescopes.

It might also be useful to organize a more official commission or working group on polarimetry; the IAU Stellar Photometry and Polarimetry Commission formally includes this topic, but has not focused much on polarimetry to date.

During the panel session, the topic of a wide-field, high-cadence polarimetric survey was brought up again. Audience members pointed out that the trend in other areas of astronomy is toward surveys of this type, and that they invariably discover new phenomena that were not previously predicted. It seems clear that such a survey is a high priority within the polarimetric community, but we will need to focus on a few compelling science cases in order to make it happen.

Overall, this group felt that in order to improve the visibility of polarimetry within the astronomical community, we need to focus on scientific results rather than the technique for its own sake. We should ask ourselves, "Why should others bother learning to do polarimetry?" and "What can polarimetry do that's unique?" Improved visualization capabilities could help significantly in this effort; displaying polarimetric results in a more attractive and intuitive way (perhaps making more use of cartoons and animations) may help us win converts. The session participants also recommended conducting an informal poll of conference attendees and other interested parties to ask two questions: "What is a really exciting science question polarimetry addresses?" and "What could you do with an all-sky high-cadence polarimetric survey?" This questionnaire is now available online at http://arwen.etsu.edu/starpol/survey.html; we encourage all interested readers to contribute their input.

## SUMMARY

Based on the results from these breakout sessions and subsequent discussion, we believe the most important "action items" for the community are as follows:

- Collect input from the polarimetry community about exciting science (existing and future) that can be used to build science cases for new instruments. The survey at http://arwen.etsu.edu/starpol/survey.html is one attempt to begin doing this.
- Establish a mailing list for the polarimetry community to improve communication and cooperation. Several informal lists exist, but we need a more centralized and better organized community.
- Establish a wiki or other website for the collection of polarimetric resources.
- Draft polarimetry-focused tutorials for undergraduate astronomy courses and training of others new to the field; look for opportunities to contribute these to existing courses or summer schools.
- When publicizing results, focus on the science more than the technique. Think carefully about visualization techniques and other ways to make your results appealing to those outside the polarimetric community.
- Continue to organize and agitate for new polarimetric instruments. Write white papers and proposals; sit on committees for new facilities. Make compelling science cases for polarimetry.
- Work toward the future goal of an all-sky, high-cadence polarimetric survey.

## ACKNOWLEDGMENTS

We thank all the conference organizers for a productive and enjoyable meeting in an ideal location. We are particularly grateful to the SOC for its encouragement of these discussions and its creation of a positive and collegial atmosphere in which to hold them. We extend sincere thanks to all the conference attendees for their participation, creativity, and collaborative spirit. JLH thanks Karen Bjorkman for initiating conversations on many of the "community priorities" topics on the first conference day.